\documentclass[a4paper,12pt,reqno,superscriptaddress,nofootinbib]{revtex4}
\usepackage[centertags]{amsmath}
\usepackage{amsfonts}
\usepackage{amssymb}
\usepackage{amsthm}
\usepackage{newlfont}
\usepackage{stmaryrd}
\usepackage{mathrsfs}
\usepackage{mathtools}
\usepackage{euscript}
\usepackage{graphicx}
\usepackage{enumerate}
\usepackage{todonotes}
\usepackage[normalem]{ulem} 
\usepackage{comment}
\usepackage{color}
\usepackage{floatrow}
\usepackage{caption}

\usepackage[draft]{hyperref}

\usepackage{tikz}
\usepackage{pgf}
\usetikzlibrary{positioning,fit,calc}
\usetikzlibrary{arrows,automata}
\usepackage{wrapfig}
\usepackage{subfigure}
\usepackage{amscd}
\usepackage{hyperref}

\theoremstyle{plain}

\theoremstyle{definition}

\theoremstyle{remark}

\newcommand{\be}{\begin{equation}}
\newcommand{\en}{\end{equation}}

\newcommand{\opunit}{\text{1}\kern-0.22em\text{l}}

\newcommand{\id}{\text{d}}
\newcommand{\dt}{\text{d}t}
\newcommand{\dx}{\text{d}x}

\DeclareMathAlphabet{\mathpzc}{OT1}{pzc}{m}{it}

\let\oldsqrt\sqrt
\def\sqrt{\mathpalette\DHLhksqrt}
\def\DHLhksqrt#1#2{%
	\setbox0=\hbox{$#1\oldsqrt{#2\,}$}\dimen0=\ht0
	\advance\dimen0-0.2\ht0
	\setbox2=\hbox{\vrule height\ht0 depth -\dimen0}%
	{\box0\lower0.4pt\box2}}

\DeclareMathAlphabet{\mathpzc}{OT1}{pzc}{m}{it}

\def\bea{\begin{eqnarray}}
\def\eea{\end{eqnarray}}
\def\ba{\begin{array}}
	\def\ea{\end{array}}

\begin{document}

\title{The asymptotic speed of reaction fronts in active reaction-diffusion systems}

\begin{abstract}
We study various combinations of active diffusion with branching, as an extension of standard reaction-diffusion processes.  We concentrate on the selection of the asymptotic wavefront speed for thermal run-and-tumble and for thermal active Brownian processes in general spatial dimensions. Comparing 1D active branching processes with a passive counterpart (which has the same effective diffusion constant and reproduction rate), we find that the active process has a smaller propagation speed. In higher dimensions, a similar comparison yields the opposite conclusion. 
\end{abstract}

\author{Thibaut Demaerel and Christian Maes \\ {\it
Instituut voor Theoretische Fysica, KU Leuven}}
\maketitle

\section{Introduction}
Speed selection in reaction-diffusion processes has been studied since more than 70 years.  The best-known example occurs in the Fisher--KPP equation, \cite{Fis,Kol},
\begin{equation}\label{kpp}
\partial_t u = D\,\partial_{xx}^2 u + \alpha\,u(1-u)
\end{equation}
for the density $u(x,t)$ of particles as function of position $x\in \mathbb{R}$ and time $t\geq 0$.  The $\alpha > 0$ couples the diffusion with the reaction part.  In the most standard set-up the process is started from $u(x,0)=1$ for $x<0$ and $u(x,0)=0$ for $x>0$. The solution of \eqref{kpp} then converges to a traveling wave $u(x,t) \to w(x-v^*t)$ where $v^*$ is the minimal speed for which there exists a traveling wave solution at all to the linearized \eqref{kpp} \cite{Ar,McK,Bram,bru,ber}. Given the context of reaction--diffusion processes however, it is natural to ask what happens when also the diffusion part of the particle motion becomes {\it active}, as was recently studied in a variety of problems such as including \cite{be,mar,pdb,ind,act,ur}.  Such studies have been made both on the mathematical and on the physical side of the question, cf. \cite{lim1,lim2,fPel,ita,men}.  The new results we add here are for thermal active processes, for run-and-tumble particles in one dimension, and for active Brownians in all dimensions.  We compare the asymptotic speed with that for passive reaction--diffusion processes with the same effective diffusion constant. For 1-dimensional processes we find that the asymptotic speed is generally lower, except when the dynamics becomes Gaussian as in activated Ornstein-Uhlenbeck processes. On the other hand, in higher dimensions (a case pioneered in \cite{Multi}) the speed turns out to be higher. Such results are quite general as can be seen also in the cases studied e.g. in \cite{ita,men}.\\  

The combination of active diffusion and chemistry can be done in many ways.  Purely on the level of differential equations for the density profile we 
could replace the (passive) diffusion part in \eqref{kpp} with its {\it active} counterpart as for example from the telegraph equation,
\begin{equation} \label{ter}
\frac 1{2\varepsilon}\,\partial^2_{tt} u +  \partial_t u = D\,\partial^2_{xx} u+ \alpha\,u\,(1-u)
\end{equation}
The extra parameter $\varepsilon$ is an inverse persistence time.
When the persistence time goes to zero, $\varepsilon\uparrow \infty$  keeping $\alpha$ and $D$ fixed, we are back to the case of \eqref{kpp}.  There is indeed bio-mathematical work done on this equation; see e.g. \cite{Kac}.
However, \eqref{ter} is not the equation one naturally obtains when starting from a biophysical process on the more mesoscopic level which motivates the problem originally.  The reason is that for the active processes we have in mind here, the particles are carrying internal degrees of freedom, like a spin or a rotation angle etc.  Only jointly, taking position and internal degrees of freedom (extra dimensions) together, is the dynamics autonomous in terms of a Markov process or a first order in time Smoluchowski-Fokker-Planck equation.   When integrating out the extra degrees of freedom, the additive structure between diffusion and reaction may disappear. To illustrate  the typical scenario, we take a binary spin degree $\sigma=\pm 1$ for simplicity.  The time-dependent density of noninteracting particles at $x$ with spin $\sigma$ is denoted by $\rho_\sigma(x,t)$.  The evolution of the joint particle density $(\rho_+(x,t),\,\rho_-(x,t))$  is
\begin{eqnarray} \label{dyn}
(\partial_t \rho_+)(x,t) &=& -c\partial_x\rho_+(x,t) + \epsilon\,(\rho_-(x,t)-\rho_+(x,t)) + \alpha\rho(x,t)\nonumber\\
(\partial_t \rho_-)(x,t) &=& c\,\partial_x\rho_-(x,t) + \epsilon\,(\rho_+(x,t)-\rho_-(x,t)) + \alpha\rho(x,t)
\end{eqnarray}
In the right-hand of \eqref{dyn}, the first term is the ballistic transport with driving speed $c$ and the second term refers to the flipping of the spin at rate $\epsilon$. The last term in each of the two equations is a pure birth (no death) at rate $\alpha$ out of the total local density $\rho(x,t) = \rho_+(x,t)+\rho_-(x,t)$.\\  
By summing and subtracting the two equations and taking another time-derivative we readily obtain the closed equation,
\begin{equation} \label{dyns}
(\partial^2_{tt} \rho)(x,t) + 2(\epsilon-\alpha)\,\partial_t \rho(x,t) = c^2\partial^2_x\rho(x,t) + 4\epsilon\,\alpha\,\rho(x,t)
\end{equation}
Comparing \eqref{dyns} with \eqref{ter}, we see first that -unsurprisingly- the chemistry in the latter equation lacks a non-linear contribution. More importantly however, we see that the chemistry of the initial model \eqref{dyns} (births with rate $\alpha$) influences the effective persistence in \eqref{dyns}.  For small persistence, $\epsilon\uparrow \infty$, and with $c^2 = 2\epsilon \,D$  we get back the linearized F--KPP equation \eqref{kpp} with birth rate $2\alpha$.  For large persistence however we see that the prefactor $\epsilon-\alpha$ becomes negative.  Interactions between particles and more chemistry can even change the equations more drastically (but we will not pursue giving further examples).\\\\
With $\alpha=0$ the evolution \eqref{dyn} describes active diffusion with effective diffusion constant
\[D_{\text{eff}}:=\lim_{t \to \infty}\frac{\langle (x_t-x_0)^2\rangle}{2t}=\frac{c^2}{2\varepsilon}
\]
With $\alpha>0$ there is a large class of positive traveling wave solutions $\rho(x,t) = w(x-vt)=e^{\lambda(v)(x-ct)}$, where the infimum for the possible values of $v$ is found to be
\[
v_{\min} =
\begin{cases}
\frac{2c}{\alpha+\varepsilon}\,\sqrt{\alpha\,\varepsilon} & \text{ when }\alpha \leq \varepsilon\\
c & \text{ when }\alpha > \varepsilon
\end{cases}
\]
Notice then that 
\begin{equation}\label{ineq'}v_{\min} < 2\sqrt{D_{\text{eff}}\alpha}\end{equation}
which (by virtue of the F-KPP-result where $v_{\text{FKPP}}=2\sqrt{D\, \alpha}$) means that the active diffusion with reproduction has a smaller overall propagation speed than the corresponding passive diffusion with diffusion constant $D=D_{\text{eff}}$; see e.g. \cite{act}.\\
We will see that this inequality remains true for thermal run-and-tumble processes in one dimension (Section \ref{trt}), but the inequality gets reversed for active Brownian motion in dimension $d\geq 2$ as shown in Section \ref{abpo}.  On the other hand the equality always holds for Gaussian processes, as can be verified readily for example for Ornstein-Uhlenbeck-activated particles that branch at a rate $\alpha$ and whose offspring inherit the same position and velocity as the parent at the time of birth. The global branch density $u(x,v,t)$ then evolves according to
\[
\partial_t u = - v \partial_x u + D_x \partial_x^2 u + \partial_v\left(\gamma vu + D_v\partial_v u\right)+\alpha u
\]
The solutions of this dynamics are of the form
\[
\overline{u}(x,t)=\frac{1}{\sqrt{2\pi \sigma^2(t)}} \exp\left(-\frac{x^2}{2\sigma^2(t)}\right)\exp(\alpha t)
\]
for some time-dependent variance $\sigma^2(t)$.
Here we look at the wave-front position $x_C(t)$ defined by
\[\overline{u}(x_C(t),t)=C\]
It shows an asymptotic velocity
\begin{equation}\label{gaga}
v_{\infty}=\lim_{t \to \infty} \frac{|x_C(t)|}{t} = 2\sqrt{D_\text{eff}\,\alpha}
\end{equation}
We start in the next Section with general ideas about speed selection including the issue of linearization and of taking different definitions for the speed, as we encountered already above.  Sections \ref{trt}--\ref{abpo} give explicit results about the minimal speed of the traveling wave for linearized active reaction-diffusion models with small driving speed.

\section{Heuristics about extracting the asymptotic speed: Branched Brownian motion}
For understanding speed selection in reaction-diffusion processes, we can rely on various definitions.  As mentioned under Eq. \eqref{kpp} we are interested in the speed of a moving front of newborns, but there are different options that nevertheless often coincide to leading  order in time.  To be more specific we remind us of a well-studied case; many things remain formally unchanged from then on.  E.g. the nonlinearity in the equation \eqref{kpp}  (and others to come) may be due to applying saturation or interaction effects; yet, the density of newborns at any moment remaining very small we are allowed to ignore the nonlinearity for the purpose of speed selection.\\ 

A branching Brownian particle starts alone at $x=0$ at time $t=0$.  The diffusion constant is $D$ and the reproduction rate is $\alpha$. Newborn particles are always spawned at the instantaneous position of the parent and they will diffuse and reproduce in the same way.\\
Then the global branch density $\rho$, defined by demanding that expected number of branches in the arbitrary interval $[a,b]$ equals $\int_a^b \rho(x,t)\dx$, solves
\begin{equation}\label{BVP}
\partial_t \rho = D \partial_x^2 \rho + \alpha \rho 
\end{equation}
subject to the initial condition $\rho(x,t=0)=\delta(x)$.\\
Hence,
\[
\rho(x,t)= \frac{1}{\sqrt{4\pi Dt}}e^{-\frac{x^2}{4Dt}+\alpha t}=\sqrt{\frac{\alpha}{4 \pi D}}e^{-\frac{x^2}{4Dt}+\alpha t - \frac{1}{2}\log (\alpha t)}
\]
and the density $\rho$ acquires the constant value $\sqrt{\frac{\alpha}{4 \pi D}}e^C$ along the line $t \mapsto (\pm x_C(t),t)$ where
\[ 
x_C(t)= \pm\sqrt{4\alpha D t^2-2Dt\log(\alpha t)-Ct} \underset{t \text{ large}}{\approx} 2\sqrt{\alpha D}\,t-\frac{1}{2}\sqrt{\frac{D}{\alpha}}\log(\alpha t)
\] 
It provides a first definition of a speed $x_C(t)/t$.  Yet it compares well to the speed of (random) position $x_\text{max}(t)$ of the rightmost branch; see \cite{Fis,Kol,McK,Ar,Bram}. In particular, it gives the correct asymptotic velocity
\begin{equation} \label{vel}
v_{\infty}:= \text{a.s.}\lim_{t \to \infty} \frac{x_\text{max}(t)}{t} = 2\sqrt{\alpha D}
\end{equation}
We can even correctly guess that the correction to the linear term in $t$ is logarithmic, although the pre-factor turns out to be too small by a factor of $3$ \cite{Bram}.\\
Another method how we could have extracted \eqref{vel} is by looking for a running-wave solution $\rho(x,t)=e^{-\lambda \,(x-vt)}$ to \eqref{BVP}. Plugging in, we get
the possibilities
\[ 
\lambda_{\pm}=\frac{v\pm \sqrt{v^2 - 4 \alpha D}}{2D}
\] 
If $v<2\sqrt{\alpha D}=:v_{\text{FKPP}}$, we are then unable to construct a positive traveling-wave solution. If  $v>v_{\text{FKPP}}$, there are $\lambda_{+} > \lambda_{\text{crit}}:=\frac{v_{\text{FKPP}}}{2D}=\sqrt{\frac{\alpha}{D}}$ and $\lambda_{-} < \lambda_{\text{crit}}$.  There is indeed a way to understand why in both cases, either for small $\lambda$ or for large $\lambda$ in the initial pattern $\rho(x,0)$ the traveling-wave solutions have a velocity greater than $v_{\text{FKPP}}$:
\begin{itemize}
\item Small $\lambda$: in this case the initial density profile has a heavy tail at large $x>0$. The higher velocity finds its origin in the offspring of those particles that started off in a relatively advanced position to begin with, without requiring a particularly large fraction of particles to move faster than $v_{\text{FKPP}}$.
\item Large $\lambda$: in this case the initial density profile has a heavy tail for $x<0$. The offspring there is so numerous that the exceptional particles which move forward faster than $v_{\text{FKPP}}$ are still numerous. Moreover, these pioneers enter regions which are very sparsely populated and therefore they cause a large relative increase in the density there. Together, this allows for an overall velocity greater than $v_{\text{FKPP}}$.
\end{itemize}
From that line of thinking, one could expect more generally that the minimal speed $v_{\min}$ attainable within the class of positive traveling wave solutions is always going to be larger than or equal to the propagation speed $v_{\infty}$ of the descendants of one single particle. For the simple case of passive diffusion that we just discussed, we arrive at the surprise that $v_{\min}=v_{\infty}$. We therefore still advance the hypothesis that the asymptotic velocity $v_{\infty}$ of the rightmost branch in a Markovian branching process may be found by calculating the minimal value $v_\text{min}$ for which we can find a positive running-wave solution for the global branch density $\rho$.
That will be our main method to proceed in the case of active diffusion, although for now we lack a physical derivation of that hypothesis.

\section{Thermal run-and-tumble particles}\label{trt}

To go beyond \eqref{dyn} and to add new results we consider here first the case of thermal run-and-tumble particles.  The temperature adds noise in the jump rates of the particles.\\
The state space $S$ of a single particle is $\mathbb{R}\times\{+,-\}$. The (time-dependent) probability density 
is denoted by $\rho(x,\sigma)=\rho_\sigma(x)$.
Its Fokker-Planck equation 
is given by
\begin{equation}\label{1p}
\begin{cases}
& \partial_t \rho_+=-c\partial_x \rho_++D\partial_x^2 \rho_++\epsilon(\rho_--\rho_+) \\
& \partial_t \rho_-=c\partial_x \rho_-+D\partial_x^2 \rho_-+\epsilon(\rho_+-\rho_-)
\end{cases}
\end{equation}
where $c$ is the driving speed and $\varepsilon^{-1}$ is the persistence time.  Compared to \eqref{dyn} we added a (passive) diffusion term referring to the presence of thermal fluctuations.
One easily verifies,  \cite{act},  that the effective diffusion constant $D_{\text{eff}}$ is given by
\begin{equation}\label{eff2}
D_{\text{eff}}:=\lim_{t \to \infty}\frac{\int_{-\infty}^{+\infty}(\rho_+(x,t)+\rho_-(x,t))x^2\dx}{2t}= D +\frac{c^2}{2\epsilon}
\end{equation}
When we extend the diffusion to include (Markovian) replication and termination of particles, this system of evolution equations for the global branch-density is altered to
\begin{equation}\label{sys'}
\begin{cases}
	\partial_t \rho_+=&-c\partial_x \rho_++D\partial_x^2 \rho_+ + \epsilon(\rho_--\rho_+)\\
	&+ (\alpha^+_{++}-\alpha^+_{--}-\delta)\rho_++(2\alpha^-_{++}+\alpha^-_{+-}+\alpha^-_{-+})\rho_-\\
	\partial_t \rho_-=&c\partial_x \rho_-+D\partial_x^2 \rho_- + \epsilon(\rho_+-\rho_-)\\
	&+ (\alpha^-_{--}-\alpha^-_{++}-\delta)\rho_++(2\alpha^+_{--}+\alpha^+_{+-}+\alpha^+_{-+})\rho_-
\end{cases}
\end{equation}
where we introduced the birth matrix $\alpha^{\sigma_1}_{\sigma_2\sigma_3}$: a parent with spin $\sigma_1$ gives birth to a descendant with spin $\sigma_2$; the spin of the parent after the birth becomes $\sigma_3$.  The parameter $\delta$ determines the rate at which a branch with (either) spin is terminated.

The first question is to find the speeds $v \geq 0$ for which there are running-wave solutions  
\[(\rho_+(x,t),\rho_{-}(x,t))\equiv (f_+(x-vt),\,f_-(x-vt))\]
to the equation \eqref{sys'}.  For simplicity we assume that the birth matrix enjoys the symmetry
\[
\alpha^+_{\sigma,\sigma'} = \alpha^-_{-\sigma,-\sigma'}
\]
We start by constructing the solution in its tail $x \to +\infty$ where $u_{\pm}(x,t) \to 0+$. There, one can  remove terms which are quadratic in $u_{\pm}$ (linearization), and propose a solution of the form $u_\pm(x,t)=f_\pm(x-vt)=v_{\pm}^0\exp(-\lambda (x-vt))$. {\it A priori}, such an {\it Ansatz} is indeed a solution provided $\lambda>0$ is a root of the polynomial
\begin{equation}
P_v(\lambda):= \det \begin{pmatrix}
D\lambda^2-(v-c)\lambda+ a& \qquad b\\
b & \qquad D\lambda^2-(v+c)\lambda+a
\end{pmatrix} \nonumber 
\end{equation}
where $a=-\epsilon+\alpha_{++}^+-\alpha_{--}^+-\delta,\quad b= \epsilon+\alpha_{+-}^-+\alpha_{-+}^-+2\alpha_{++}^-$.  In other words, 
\begin{equation}\label{pol}
P_v(\lambda) = (D\lambda^2-v\lambda+a)^2-(c\lambda)^2 - b^2
\end{equation}
Requiring that the coefficients $v_\pm^0$ are non-negative, yields the additional constraint
\begin{equation}\label{ineq}
D\lambda^2 -(v-c)\lambda +a\leq 0
\end{equation}
The positive roots $\lambda^*$ of $P_v$ that obey the constraint \eqref{ineq} satisfy
\begin{equation}\label{exp}
(\lambda^*)^2\,D-v\lambda^*+a=- \sqrt{b^2+(c\lambda^*)^2}
\end{equation}
\subsection{The non-thermal case: $D=0$}
When $D=0$, the condition that $P_v(X)=(v^2-c^2)X^2-2av X+a^2-b^2$ has real roots at all reqiures that its discriminant is non-negative:
\begin{equation}\label{b1}v^2 \geq \frac{b^2-a^2}{b^2}c^2
\end{equation}
On the other hand, the inequality \eqref{ineq} (which now reads $-(v-c)\lambda^* + a \leq 0$) and the requirement that $\lambda^*>0$ implies the additional requirement
\begin{eqnarray}
&& \lambda^*\geq \frac{a}{v-c}\text{ when } v> c \text{ and } a \geq 0\label{b2}\\
&& \lambda^*\leq \frac{-a}{c-v} \text{ when }v< c \text{ and } a\leq 0\label{b3}\\
&& \text{All $\lambda^*>0$ fine when $v\geq c$ and $a\leq 0$}, \nonumber\\
&& \text{otherwise no $\lambda^*>0$ satisfy the inequality.}\nonumber
\end{eqnarray}
We now consider 4 disjoint cases:
\begin{itemize}
\item $a< 0$ and $b^2-a^2> 0$: we set $v$ equal to $c\sqrt{\frac{b^2-a^2}{b^2}}$ (so that \eqref{b1} reduces to an equality) and we find that $P_v$ has a double positive root at $X=\frac{|b|}{|a|c}\sqrt{b^2-a^2}$ which is less than or equal to $\frac{|a|}{c-v}=\frac{|ab|}{c(|b|-\sqrt{b^2-a^2})}$ so that also \eqref{b3} is fulfilled. Hence $v_{\min}=c\sqrt{\frac{b^2-a^2}{b^2}}$
\item $a< 0$ and $b^2-a^2\leq 0$: For $v=0$, $P_v$ has the roots $\pm \frac{\sqrt{a^2-b^2}}{c}$ of which the greatest one is non-negative, yet smaller than $\frac{|a|}{c}=\frac{|a|}{c-v}$: hence the condition \eqref{b2} is met and we conclude that $v_{\min}=0$ in this case. \\Remember that for e.g. the model \eqref{dyn}, $a=\alpha-\varepsilon$ and $b=\alpha+\varepsilon$ so that $b^2-a^2=4\alpha\epsilon$ can never be negative.
\item $a\geq 0$: Let us put $v=c(1+\varepsilon)$ (with the purpose of letting $\epsilon \to 0+$), the largest root of $P_v$ diverges to $+\infty$ approximately as $\frac{4av}{2(v^2-c^2)}\sim \frac{a}{c\varepsilon}$. As far as the bound \eqref{ineq} is concerned, we have to meet \eqref{b1} wherein the right-hand-side $\frac{a}{v-c}= \frac{a}{c\varepsilon}$. Plugging $X=\frac{a}{c\epsilon}$ into $P_v$ yields the result $-b^2\leq 0$. Hence $P_v$ has a root $\lambda^*$ larger than $X$, so that \eqref{b1} is fulfilled. Hence $v_{\min}=\inf_{\epsilon>0}c(1+\epsilon)=c$ (although that minimal velocity may not by strictly attained in this case)
\end{itemize} 
\eqref{b1} and \eqref{b2} combined yield the result
\begin{equation}
v_{\min}=\begin{cases} 
0 & \text{ when }a<0\text{ and }b^2> a^2\\
c\sqrt{\frac{b^2-a^2}{b^2}} & \text{ when }a<0\text{ and }a^2\geq b^2\\
c & \text{ when }a\geq 0
\end{cases}
\end{equation} 
\subsection{Driving speed $c$ small}
When the driving speed $c$ is relatively small, we can proceed to make concrete estimates. 
Using $\sqrt{1+x}\geq 1+\frac{x}{2}-\frac{x^2}{8}$ in \eqref{exp}, we get
\begin{equation}\label{in}
D(\lambda^*)^2-v\lambda^*+a\leq -\left(b+\frac{(c\lambda^*)^2}{2b}-\frac{(c\lambda^*)^4}{8b^3}\right)\leq -\left(b+\frac{(c\lambda^*)^2}{2b}-\frac{c^4}{8b^3}\left(\frac{v\lambda^*-a}{D}\right)^2\right)
\end{equation}
where, to obtain the last inequality, we used \eqref{ineq} again.  Per consequence the quadratic polynomial 
\[
q(x) := \left(D+\frac{c^2}{2b}-\frac{c^4v^2}{8b^3D^2}\right)x^2-\left(1-\frac{ac^4}{4b^3D^2}\right)vx+a+b-\frac{a^2c^4}{8b^3 D^2}
\]
diverges to $+\infty$ asymptotically in $\lambda^*$ and still, from \eqref{in} acquires a negative or zero value in a bounded interval. Hence, the discriminant of this polynomial has to be non-negative, which implies
\[
v^2 \geq 4\left(D+\frac{c^2}{2b}\right)(a+b)+O(c^4)
\] 
So the positive wave-front in this range of velocities exists and therefore the minimal speed satisfies
\begin{equation}
v_{\min}^2\leq 4\left(D+\frac{c^2}{2b}\right)(a+b)
+O(c^4)\end{equation}
The important remark now is that, because $b\geq \epsilon$, we have by \eqref{eff2} that $D+\frac{c^2}{2b}\leq D_{\text{eff}}$ and hence the minimal traveling-wave speed is less than the effective F-KPP speed:
\begin{equation}\label{lele}
v_{\min}^2\leq 4D_{\text{eff}}\alpha_{\text{eff}}
\end{equation}
where $\alpha_\text{eff} = 
\alpha_{++}^{+}+\alpha_{+-}^{+}+\alpha_{-+}^{+}+\alpha_{--}^{+} - \delta = a+b$.

\section{Active Brownian particles with offspring}\label{abpo}
For two-dimensional active Brownian particles, \cite{ur}, we have two coordinates for spatial location and one angle for the orientation of the propagation speed.  In dimensions $d\geq 2$ and for independent active Brownian particles the Fokker-Planck equation for the probability density $\rho(x_1,...,x_d,\theta_1,...,\theta_{d-1},t)$ relative to the volume element $\left(\prod_{j=1}^{d-1}\dx_j\,\sin^{d-j-1}(\theta_j)\id \theta_j\right)\dx_d$ is given by
\begin{equation} \label{EOM}
	\partial_t \rho = -\mathbf{c}(\theta) \cdot \nabla_x \rho +D_x \nabla_x^2 \rho + D_{\theta} \nabla_{\theta}^2 \rho
\end{equation}
where the velocity $\mathbf{c}$ is defined in terms of the hyperspherical angles $\theta_j$ through the formula $v_j(\theta) = c \left(\prod_{k=1}^{j-1}\sin(\theta_k)\right)\cos(\theta_j)$ for $2\leq j \leq d-1$ while $c_1(\theta)=c\cos(\theta_1)$ and $c_d(\theta)=c\left(\prod_{k=1}^{d-1}\sin(\theta_k)\right)$.
We want the projection of the equation of motion on the the direction of propagation.
The marginal density 
\[\rho_m(x_1,\theta_1,t):=\int_{\mathbb{R}^{d-1}}\left(\prod_{j=2}^d\dx_j\right) \int_{(0,\pi)^{d-2}\times (0,2\pi)}\left(\prod_{j=2}^{d-1}\sin^{d-j-1}(\theta_j)\id \theta_j\right)\rho(x,\theta,t)\]
can easily be shown to evolve according to the following much simpler evolution equation
\begin{equation}\label{FP}
\partial_t \rho_m = -c\cos(\theta_1)\frac{\partial \rho}{\partial x_1}+D_x\frac{\partial^2 \rho}{\partial x_1^2} + D_{\theta}\sin^{2-d}(\theta_1) \frac{\partial}{\partial \theta_1}\left(\sin^{d-2}(\theta_1)\frac{\partial \rho}{\partial \theta_1}\right)
\end{equation}

\subsection{Effective diffusion constant}
Multiplying \eqref{FP} with $x_1^2$ resp. $x_1 \cos(\theta_1)$ and applying the integration $\int_{-\infty}^{+\infty} \dx_1 \int_{0}^{\pi}\sin^{d-2}(\theta_1)\id \theta_1$ yields the equations
\begin{equation}\label{sys}
\begin{cases}
& \frac{\id}{\dt}\langle x_1^2\rangle_t = 2c\langle x_1 \cos(\theta_1)\rangle_t + 2D_x \\
& \frac{\id}{\dt}\langle x_1\cos(\theta)\rangle_t = c\langle \cos^2(\theta_1)\rangle_t - (d-1)D_\theta \langle x_1 \cos(\theta_1)\rangle_t \\
& \frac{\id}{\dt}\langle \cos^2(\theta_1)\rangle_t=2D_{\theta}\left(1-d\langle\cos^2(\theta_1)\rangle_t\right)
\end{cases}
\end{equation}
From the last equation, $\langle \cos^2 (\theta_1)\rangle_t \to \frac{1}{d}=\frac{\int_0^\pi\sin^{d-2}(\theta_1)\cos^2(\theta_1)\id \theta_1}{\int_0^\pi \sin^{d-2}(\theta_1)\id \theta_1}$ as $t \to \infty$. Inserting that in the second equation of \eqref{sys}, we see that $\langle x_1 \cos(\theta_1) \rangle_t \to \lim_{t \to \infty}\frac{c}{D_{\theta}(d-1)}\langle \cos^2 (\theta_1)\rangle_t = \frac{c}{D_{\theta}}\frac{1}{d(d-1)}$. Therefore, the first equation of \eqref{sys} finally implies that $\langle x^2 \rangle_t \to 2(D_x + \frac{c^2}{d(d-1)D_{\theta}})$. The effective diffusion constant is therefore given by
\begin{equation}\label{eff}
D_{\text{eff}}:=\lim_{t \to \infty} \frac{\langle x_1^2 \rangle_t}{2t}=D_x + \frac{c^2}{d(d-1)D_{\theta}}=\lim_{t \to \infty} \frac{\left\langle \sum_{j=1}^dx_j^2 \right\rangle_t}{2d\,t}
\end{equation}

\subsection{Propagation speed of front of newborn particles}\label{abpspeed}
We add a growth-term $\alpha \rho$ to \eqref{EOM} to describe the global density for active Brownians with branching at rate $\alpha$.  That addition corresponds to an extra term $\alpha \rho_m$ in \eqref{FP}:
\begin{equation} \label{EOM'}
\partial_t \rho_m = -c\cos(\theta)\partial_x \rho_m +D_{x}\partial_x^2\rho_m+ D_\theta\sin^{2-d}(\theta)\partial_{\theta}\left(\sin^{d-2}(\theta)\partial_{\theta}\rho\right)+\alpha \rho_m
\end{equation}
New particles inherit the orientation $\theta$ of the parent-particle. To estimate the velocity of the newborns, we need the \textit{positive} traveling-wave solution $\rho_m(x,\theta,t)=e^{-\lambda(x-vt)}f(\theta)$ (writing $x,\theta$ instead of $x_1,\theta_1$) to \eqref{EOM'}.  Plugging in the traveling-wave Ansatz, we get
\begin{eqnarray}
&& 0 = [\alpha+(c\cos(\theta)-v)\lambda  + D_x \lambda^2]f(\theta) + D_\theta\sin^{2-d}(\theta)\left(\sin^{d-2}(\theta)f'(\theta)\right)'\nonumber\\
&&=:(\alpha-v\lambda+D_x\lambda^2)f(\theta)-(Lf)(\theta)\label{eq}\label{kapp}
\end{eqnarray}
where the differential operator $L$ is Sturm-Liouville for the desired boundary condition $f'(0)=f'(\pi)=0$ and in the Hilbert space defined by the inner product
\[\langle g,h\rangle=\int_0^\pi \id \theta \sin^{d-2}(\theta)\,\overline{g(\theta)}\,h(\theta)\]
Equation \eqref{eq} forces us to find a positive function $f$ that is an eigenvector of $L$. A standard result of Sturm-Liouville theory confirms that there is precisely one such eigenvector and it is the one corresponding to the lowest eigenvalue $\kappa$ of $L$ and is the infimum of the Rayleigh-Ritz quotient:
\begin{eqnarray}
Q(f)&:=&\frac{\langle f,Lf\rangle}{\langle f,f\rangle}=\frac{\int_0^\pi \id \theta \,\sin^{d-1}(\theta)\left(-c\lambda \cos(\theta)|f(\theta)|^2+D_{\theta}|f'(\theta)|^2\right)}{\int_0^\pi \id \theta \, \sin^{d-1}(\theta)|f(\theta)|^2}\nonumber\\
\kappa &=& \inf_{f>0} Q(f) = \alpha-v\lambda+D_x\lambda^2 \label{try}
\end{eqnarray}
where the fact that $\kappa=\alpha-v\lambda+D_x\lambda^2$ follows from \eqref{kapp}.\\
Plugging the trial function $\tilde{f} \equiv1$ into the Rayleigh-Ritz quotient yields the bound $\kappa\leq 0$. The equation \eqref{try} then reduces to the inequality $0\leq \alpha-v\lambda+D_x\lambda^2$ which requires a non-negative discriminant, or equivalently $v\geq 2\sqrt{\alpha D_x}$. Therefore also $v_{\min}\geq 2\sqrt{\alpha D_x}$: the active component of the particle diffusion certainly speeds up the wave relative to the case where the driving $c$ is set to zero.\\
The more interesting question however is whether $v_{\min} \geq 2\sqrt{\alpha D_{\text{eff}}}=2\sqrt{\alpha(D_x + \frac{1}{D_{\theta}}\frac{c^2}{d(d-1)})}$. That is shown to hold in the Appendix, at least for small $c$, by plugging better trial functions $\tilde{f}$ in the Rayleigh-Ritz quotient (by ``better", we mean that $\tilde{f}$ has to better approximate the actual leading eigenvector of $L$). We calculate there that, up to order $c^4$,
\begin{equation}
\label{sho}
v_{\min} = 4\alpha D_{\text{eff}}+\frac{\alpha^2c^4}{d^3(d-1)^3(d+2)D_{\theta}^3D_{\text{eff}}}
\end{equation} 
In other words, at least for sufficiently small driving $c$, the minimal speed $v_{\min}$ exceeds the effective F-KPP speed, in contrast with the situation for run-and-tumble processes in $d=1$; see \eqref{lele}.
In high dimensions $d\uparrow \infty$ the Gaussian equality \eqref{gaga} is restored.

\section{Conclusions}
The present work is adding physically relevant complications to an old problem.  The reaction-diffusion equations now contain both thermal and nonthermal noise, making them more interesting in view of their application to active particles in a thermal environment. The main goal was to collect information on the asymptotic speed of the traveling wave front of newborns provided the initial population is sufficiently compactly supported. Our method rests on the assumption that the relevant speed is always equal to the minimal speed that can be attained in a traveling wave solution to the linearized dynamics. Using this assumption and method, we found evidence that the front slows down with respect to a corresponding passive case in one-dimensional run-and-tumble processes while the opposite is true for active Brownians in dimension $d>1$. We have found explicit expressions detailing the speed's dependence on the various relevant parameters such as persistence, driving speeds, dimension and temperature. Such studies can be extended in various directions.  First there is the obvious question of understanding corrections to the asymptotic speed such as via logarithmic and other corrections in time, \cite{bru,Bram}.  Secondly there is the wider study of surface growth, interface motion and other growth processes in the presence of active components. To the best of our knwoledgde the case of active Eden of active DLA has not been studied yet. Such active interface models would be relevant, so it seems, within biological contexts of growth. Fluctuations around the speed could then be compared with those for other front motions such as studied e.g. in the Kardar-Parisi-Zhang universality class.\vspace{2cm}

\noindent {\bf Acknowledgment:}
We thank Pierre de Buyl, Bernard Derrida and Pierre Gaspard for encouraging remarks and discussions.

\appendix

\section{Asymptotic speed for thermal active Brownians}\label{appe}
We give the derivation of \eqref{sho}.  If we write $f=e^g$ ($f$ was already required to be a positive function) in \eqref{try}, we have
\begin{eqnarray}
&& c\lambda \cos(\theta)f(\theta) + D_\theta\sin^{2-d}(\theta)\left(\sin^{d-2}(\theta)f'(\theta)\right)' \equiv Cf(\theta) \text{ and } f'(0)=f'(\pi)=0\nonumber\\
&&\Leftrightarrow c\lambda\cos(\theta)+D_{\theta}\left(\sin^{2-d}(\theta)\left(\sin^{d-2}(\theta)g'(\theta)\right)'+g'(\theta)^2\right)\equiv C \text{ and } g'(0)=g'(\pi)=0\label{log}
\end{eqnarray}
We solve \eqref{log} by iteration until we are close enough to the actual solution. For this, write $h=g'$, decompose into $h=h_0+h_1+h_2$ and let $h_{0,1,2}$ solve
\begin{equation}\label{i}
\begin{cases}
\sin^{2-d}(\theta)\left(\sin^{d-2}(\theta)h_0(\theta)\right)'+\underbrace{\frac{c\lambda}{D_\theta}}_{=:\xi} \cos(\theta)=0&  h_0(0)=0\\
\sin^{2-d}(\theta)\left(\sin^{d-2}(\theta)h_1(\theta)\right)'+h_0(\theta)^2=C_0&  h_1(0)=0\\
\sin^{2-d}(\theta)\left(\sin^{d-2}(\theta)h_2(\theta)\right)'+h_1(\theta)^2+2h_1(\theta)h_0(\theta)=C_1& h_2(0)=0 \\
C_0 =\int_0^\pi \id \theta\, \sin^{d-2}(\theta)h_0^2(\theta)/\int_0^\pi \id \theta\, \sin^{d-2}(\theta)& \\
C_1 =\int_0^\pi \id \theta\, \sin^{d-2}(\theta)\left(h_1^2(\theta)+2h_1(\theta)h_0(\theta)\right)/\int_0^\pi \id \theta\, \sin^{d-2}(\theta)&
\end{cases}
\end{equation}
Then by construction, $f=e^{\int h}$ solves
\begin{equation}\label{plug}\frac{-Lf}{f} = D_\theta \left(C_1+C_2+\underbrace{h_2(2h_0+2h_1+h_2)}_{=\text{small, i.e. $O(\xi^5)$ (see later)}}\right)\end{equation}
The solution to \eqref{i} is given by
\begin{equation}
\begin{cases}
h_0(\theta)= \frac{\xi}{d-1}\sin(\theta)&  C_0=\frac{\xi^2}{d(d-1)}\\
h_1(\theta)=\frac{\xi^2}{d(d-1)^2}\sin(\theta)\cos(\theta)&  C_1=\frac{\xi^4}{d^3(d-1)^3(d+2)}\\
h_2(\theta)=O(\xi^4)& 
\end{cases}
\end{equation}
Inserting \eqref{plug} into the Rayleigh-Ritz quotient then yields a bound of the form
\begin{equation}\label{est}
\kappa^* \leq -D_{\theta}\left(\frac{\xi^2}{d(d-1)}+\frac{\xi^4}{d^3(d-1)^3(d+2)}\right)+M\xi^5
\end{equation}
where the constant $M$ can be estimated explicitly if the need would arise. Plugging \eqref{est} into \eqref{eq} yields
\begin{equation}
0 \leq \alpha -v\lambda + \underbrace{\left(D_x + \frac{c^2}{d(d-1)D_{\theta}}\right)}_{=D_{\text{eff}}}\lambda^2 + \frac{c^4}{d^3(d-1)^3(d+2)D_{\theta}^3} \lambda^4 + M \frac{c^5}{D_{\theta}^4}\lambda^5=:p_v(\lambda)
\end{equation}
We can scan $p_v$ along $\lambda=\frac{v}{2D_{\text{eff}}}$ (which is the minimum point of the parabolic part of $p_v$):
\begin{equation}
p_v(v/(2D_{\text{eff}})) = \alpha - \frac{v^2}{4D_{\text{eff}}}+\frac{c^4v^4}{16d^3(d-1)^3(d+2)D_{\theta}^3D_{\text{eff}}^4} + M \frac{c^5v^5}{32D_{\theta}^4D_{\text{eff}}^5}
\end{equation}
The smallest root $v_0$ of this equation (which is a lower bound for $v_{\min}$) is
\begin{equation}\label{res}
v_0^2 = 4\alpha D_{\text{eff}}+\frac{\alpha^2c^4}{d^3(d-1)^3(d+2)D_{\theta}^3D_{\text{eff}}} + O\left(M\frac{\alpha^2c^4}{D_{\theta}^3 D_{\text{eff}}}\frac{c}{\sqrt{\alpha D_{\text{eff}}}}\right)
\end{equation}
One can in fact prove, by showing that the iteration \eqref{i} will in fact converge to the actual solution, that $v_{\min}$ agrees with $v_0$ to fourth order in $c$ in the sense of the expression \eqref{res}.

\end{document}